\documentstyle[11pt,aaspp4]{article}

\received{}
\revised{}
\accepted{}
\journalid{}{}
\articleid{}{}
\paperid{}
\cpright{}{}
\ccc{}

\slugcomment{Submitted to the Astrophysical Journal}

\lefthead{Stassun \& Wood}
\righthead{Classical T Tauri Stars}

\begin{document}

\title{Magnetic Accretion and Photopolarimetric Variability \\
in Classical T Tauri Stars}

\author{Keivan Stassun\altaffilmark{1} and Kenneth Wood\altaffilmark{2}}

\altaffiltext{1}{Astronomy Dept., University of Wisconsin, 475 N.
Charter St., Madison, WI 53706; keivan@astro.wisc.edu}

\altaffiltext{2}{Smithsonian Astrophysical Observatory,
60 Garden Street, Cambridge, MA 02138; kenny@claymore.harvard.edu}

\authoremail{keivan@astro.wisc.edu}

\begin{abstract}
We employ a Monte Carlo radiation transfer code to investigate the
multi-wavelength photopolarimetric variability arising
from a spotted T Tauri star surrounded by a dusty circumstellar disk.
Our aim is to assess the ability of the magnetic accretion model
to explain the observed photopolarimetric variability of classical T Tauri stars,
and to identify potentially useful observational diagnostics of T Tauri
star/disk/spot parameters. We model a range of spot sizes, spot latitudes,
inner disk truncation radii, and system inclination angles, as well as multiple
disk and spot geometries. We find that the amplitude, morphology, and wavelength
dependence of the photopolarimetric variability predicted by our models are generally 
consistent with existing observations; a 
flared disk geometry is required to reproduce the largest observed polarization 
levels and variations. Our models can further explain stochastic
polarimetric variability if unsteady accretion is invoked, in which case irregular ---
but correlated --- photometric variability is predicted, in agreement with
observations. 

We find that variability in percent polarization 
is by itself an unreliable diagnostic as certain system geometries do not
produce any variability in linear polarization (contrary to the commonly held
notion that hot
spots will necessarily produce periodic polarimetric variability).  
Observations of variability in polarization position angle, however, could provide 
useful constraints on system inclination. The observation of wavelength-dependent
polarization position angles, attributed by some to interstellar effects, is 
naturally explained by our models.
Certain system geometries yield peculiar photometric light curve morphologies, 
the observation of which could also serve to constrain system inclination. We do
not find useful diagnostics of disk truncation radius, nor do we find significant
differences when we model the spots as rings.

We also investigate the
reliability of modeling spot parameters via analytic fits to multi-band
photometric variations. We
find that commonly used analytic models consistently recover input model
parameters
but that inferred spot temperatures are more sensitive to uncertainties
in the photometric data than previous modeling would suggest.
\end{abstract}

\keywords{accretion, accretion disks --- radiative transfer ---
polarization --- stars: pre-main sequence --- stars: rotation --- stars: spots}

\section{Introduction}

T Tauri stars are low-mass pre-main sequence (PMS) stars with a rich
history of observational and theoretical study, dating back to Joy's (1945)
discovery paper. Seen initially as stellar
oddities, T Tauri stars have become paradigmatic of early stellar evolution.
The data
paint an intricate picture of stars in their earliest evolutionary stages:
Excess infrared and sub-millimeter emission indicate
extended circumstellar disks of gas and dust; excess ultraviolet emission,
hydrogen emission lines, and erratic brightness variations suggest hot
accretion zones, where infalling material impacts the stellar surface;
photospheric spots, similar in nature to sunspots, reveal magnetic activity.

Despite this wealth of information regarding the nature of T Tauri stars,
they remain
somewhat enigmatic objects. In particular, our understanding of the accretion
process is at present incomplete, and is currently a subject of intense study.
Until recently, the commonly accepted mechanism for accretion of circumstellar
material onto stellar surfaces was via a ``boundary layer" model
(e.g. \cite{basri89}),
in which disk material falls onto the star along the midplane, producing an
equatorial ring of hot ($\sim 10,000$ K), optically thin material.
More recently, however, attention has turned to an accretion model which
attempts to account for the influence of the stellar magnetic field, and
which serves to
explain such phenomena as the observed bimodal rotation period distribution
(e.g. \cite{aherbst92}), inverse P Cygni line profiles
(\cite{hart94}; \cite{edwards94}), and the presence of photospheric
hot spots (\cite{bouvier93}; \cite{herbst94}).

The magnetic accretion model (Ghosh \& Lamb 1979a, 1979b; \cite{konigl};
\cite{shu94}; \cite{ostrikershu}; \cite{najita95}) as applied to PMS stars is
characterized by: a star
surrounded by a geometrically thin, optically thick, circumstellar disk which
is truncated within a certain inner radius; a
stellar dipole magnetic field which threads the disk; accretion ``streams" of 
disk material which flow along field lines and impact the star at the magnetic 
poles, producing hot spots there
(or ``rings"; \cite{mahdavi}). The presence of hot spots or rings on
the stellar
surface has observable consequences. Periodic photometric variations will be
produced as the star's rotation brings the spots into and out of view;
periodic polarimetric variations have also been predicted (e.g. 
\cite{gullgahm}) due to periodically varying illumination of the disk by the spots.

Photometric variability is a ubiquitous characteristic of
T Tauri stars. And {\it periodic} photometric variability has now been confirmed
in hundreds of systems, in star-forming regions throughout the Galaxy:
Bouvier et al. (1993) in their study of 24 T~Tauri stars in Taurus-Auriga
found evidence for periodic variability in at least 20; in Lupus, Wichmann
et al. (1998) found periodic variability in 34 of 46 targets studied;
in Orion, Herbst and
collaborators have to date identified 75 periodic variables
(\cite{choi}); in NGC 2264, Makidon et al. (1997) find some 200
periodic variables. The periods of the observed variability
range from about one day to several tens of days, with the short period limit
likely determined by the sampling of the observing programs.
In most systems the spots persist stably over multiple rotation periods and, in some
cases, over multiple observing seasons. A favored explanation for such periodic
variability is stellar spots; periodic photometric variations have
been modeled analytically in numerous systems by several authors (e.g.
\cite{bouvier93}) within this context in order to derive spot properties
(size, temperature, position). Taken at face value, the results of such
modeling suggest that spots on weak-lined T Tauri systems tend to be cool
(cooler than the photosphere) while classical T Tauri systems  can possess both
hot and cool spots. In some systems, hot and cool spots have been found
concurrently (e.g. \cite{vrba86}). Though hot accretion spots seem to manifest
themselves most often as irregular variables, stable photometric variability
due to hot spots has been seen in numerous systems (e.g. \cite{bouvier93}). 
It is these systems which we model in this study.

Polarimetric variability has also been detected in numerous systems (e.g.
\cite{drissen89}),
though so far {\it periodic} polarimetric variability has not been found\footnote{
Periodic polarimetric variability has been reported for RY Lup (see references in
\cite{drissen89}) but we have been unsuccessful in tracing original publication of this
result.}. As a result, alternative mechanisms
for the observed variability --- such as obscuration by circumstellar
material --- have been proposed (e.g. \cite{gullgahm}). More typically,
erratic polarimetric variability  of $\sim 0.5\%$ has been observed 
with only a few
systems exhibiting variability of 2-3\% (\cite{gullgahm}). Numerous systems have 
exhibited large position angle variations as well.
In a few cases, photometric and polarimetric
variability have been observed to be anti-correlated (\cite{drissen89}; \cite{gullgahm}), 
in the sense of lower polarization at maximum light.

Detailed calculations of spot/ring geometries have been carried out 
(\cite{mahdavi}) as well as
Monte Carlo simulations of photopolarimetric variability of a spotted star
plus flat disk (\cite{kenny96}) where predictions were made of the
$V$ band photopolarimetric variability for one particular star-spot-disk
geometry.
In this paper, we extend the work of Wood et al., and perform simulations of
the photopolarimetric variability at several wavelengths ($U$, $V$, $I$, and
$K$) and for a large grid of spot latitude, spot size, and system inclination geometries.
We also investigate the effects of a flared disk, and of ring-like spots, on
the resulting
photopolarimetric variability. As the magnetic accretion paradigm gains 
popularity as a generalized model for accretion in T Tauri stars, it is timely to
confront the model with the existing data and to seek out robust observational 
diagnostics of accretion signatures in these systems. Thus our aim here is to 
assess the
ability of the magnetic accretion model to explain existing observations of 
photopolarimetric variability in classical T Tauri stars, to
identify observational diagnostics
which may prove useful in constraining T Tauri system spot/disk properties of
interest, and to make predictions of unusual or otherwise distinctive
observable phenomena arising from the magnetic accretion model. 
Basic physical properties 
of spots such as temperature and size are particularly germane to this discussion, 
and so
we examine the validity of an analytic inversion technique commonly used to infer
spot properties from photometric variability by inverting our simulated 
data to retrieve known spot temperatures and sizes.

In Section 2 we describe the Monte Carlo radiation transfer technique
employed. The results of our simulations are presented in Section 3, wherein we
discuss potentially useful photopolarimetric diagnostics of T Tauri system
parameters. We compare our models
to observations in Section 4, and generally discuss the ability of the magnetic
accretion model to explain the existing photopolarimetric data. In Section 5 we 
examine an analytic
inversion technique commonly used to model observations of photometric
variability. Lastly, we summarize our findings in Section 6.

\section{Model Parameters and Radiation Transfer Technique}

The models that we investigate comprise a central spherical star of
temperature $T_\star$, with two
diametrically opposed circular spots with radii $\theta_s$, located at
latitudes $\pm\phi_s$, both with temperature $T_s$.
This spotted star is surrounded by a
dusty equatorial circumstellar disk with inner and outer radii $R_{\rm in}$
and $R_{\rm max}$.  We investigate both flat and flared disk geometries.
The radiation transfer is performed with the Monte Carlo technique.

\subsection{Stellar Parameters}

For all our simulations we keep the stellar and spot temperatures fixed at
$T_\star = 4000$K and $T_s = 10000$K. These parameters are consistent with
those typically quoted for classical T Tauri stars; T Tauri stars are
typically of spectral type early K and later, and detailed spectroscopic
studies of classical T Tauri stars typically reveal featureless, hot ``veiling"
components which are bluer than photospheric colors (e.g. Basri \& Batalha
1990).
The parameters that we vary are
the spots' angular size on the star, $\theta_s$, and the latitude of the
diametrically opposed spots,
$\pm\phi_s$.  We investigate spots with covering factors of 6\%, 2\%, and
.07\% of a hemisphere ($\theta_s = 20^\circ$, $11.5^\circ$, and $6.6^\circ$,
respectively), and choose spot latitudes in the range
$45^\circ<\phi_s<80^\circ$, thus simulating low and high latitude spots.

Strictly speaking, theory predicts that the geometry of accretion streams
result in hot accretion ``rings" on stellar surfaces (\cite{mahdavi};
\cite{kenyon94}), as opposed to
the simple ``spot" accretion geometry we have modeled.
Our Monte Carlo technique is capable of modeling these more complex
geometries if desired,
but we find that modeling the accretion geometry as rings
as opposed to spots does not have an observable effect on the photopolarimetry
arising from T Tauri systems (Section~3.3). Thus, for the sake of simplicity,
and for computational ease, we model the accretion geometry as simple spots
with parameters $\phi_s$ and $\theta_s$.

\subsection{Disk Geometry}

We consider two geometries for the circumstellar disk.  The first is the
flat disk used by Wood et al. (1996).  The geometry is given by
\begin{equation}
\rho={{\rho_0}\over{x^2+y^2+z^2/b^2}}\; .
\end{equation}
This particular form for the circumstellar geometry enables the Monte Carlo
radiation transfer to proceed rapidly since the randomly chosen scattering and
absorbing locations are found analytically and numerical integration is not
required.  We choose $b=10^{-3}$ which yields a
very flat disk-like structure and set the equatorial optical depth
$\tau_{\rm eq} = 10^3$.  We have kept the equatorial optical depth fixed for
all wavelengths and only change the dust scattering parameters with
wavelength (see Section~2.3).  For the more realistic case of a flared disk
geometry we set the disk optical depth according to the wavelength dependent
opacity.

The second geometry we consider is a flared disk as prescribed by Shakura
\& Sunyaev (1973),
\begin{equation}
\rho=\rho_0\exp \{-{1\over 2}[z/(h(\varpi)]^2 \}/\varpi^\alpha \; 
\end{equation}
where $\varpi$ is the radial coordinate in the disk midplane. The scale 
height increases with radius according to
\begin{equation}
h=h_0\left({{\varpi}\over{R_\star}}\right)^\beta \; .
\end{equation}
The degree of flaring, $\beta$, lies in the range $9/8 < \beta < 5/4$
(\cite{kh87}), and the scale height at the stellar surface is
typically taken to be $h_0=0.03R_\star$.  For our simulations we adopt 
$\alpha = 15/8$, $\beta = 9/8$, $h_0 = 0.025 R_\star$, and a disk mass of
$M_{\rm disk} = 1 \times 10^{-3} M_\odot$.
For both flat and flared disks we set the outer disk radius to be
$R_{\rm max} = 100$AU, and we vary the disk truncation radius, $R_{\rm in}$
from $R_\star$ (no hole) to $15R_\star$, thus covering the range of values
generally quoted for inner disk holes (e.g. Beckwith et al. 1990; 
Kenyon, Yi, \& Hartmann 1996).

\subsection{Dust Opacity and Scattering Parameters}

The circumstellar dust is assumed to have the same properties as a
Mathis, Rumpl, \& Nordsieck (1977) interstellar dust mixture with a total
opacity, $\kappa$, and albedo, $a$.  The scattering phase
function is modeled by a Heyney-Greenstein phase function with asymmetry
parameter, $g$, and polarization, $p$.  The variation of the dust parameters
with wavelength is presented in Table~1.  It is possible that the
circumstellar dust properties we adopt are different from interstellar dust.
However,
Whitney, Kenyon, \& G\'{o}mez (1997) found that these parameters provide a good match to the
colors of Taurus protostars and we adopt them for our simulations also
(though our Monte Carlo code could use any dust mixture just as easily).

\subsection{Radiation Transfer}

The Monte Carlo radiation transfer technique has been used in previous
investigations of dust scattering and absorption in the circumstellar
environments of T~Tauri stars (e.g., Whitney \& Hartmann 1992, 1993;
Whitney et al. 1997; Fischer, Henning, \& Yorke 1994; Bastien \&
Menard 1988). Our code simulates the transfer of ``photon packets''
though the circumstellar disk with the output for the present investigation
being the spatially unresolved flux and polarization.  Since the disk is 
illuminated by
an asymmetric source (spotted star), when the photons exit the system we
place them into direction-of-flight 
bins that are determined from the
photon's latitudinal and azimuthal directions (\cite{kenny96}). 
The stellar photons are released from the star and spots 
in proportion to the relative luminosities and surface areas of the star
and spots. We assume that the luminosities of the star and spots are Planck
functions, $B_\lambda$.  Therefore, the relative number of photons released from
the spot and star at a given wavelength is
\begin{equation}
{{N_s}\over{N_\star}}={(1-\cos\theta_s)\over (\cos\theta_s)}
{{B_\lambda(T_s)}\over{B_\lambda(T_\star)}}\; 
\end{equation}
where again $\theta_s$ is the spot angular radius. 
The different wavelengths that we investigate are therefore simulated by
releasing spot and stellar photons according to this equation and the opacity
and scattering parameters are those presented in Table~1.  In our
simulations we assume that all the photons originate from the star and neglect
any disk emission.  This is a reasonable assumption at optical wavelengths.
However, for systems in which the disk extends to the stellar surface we
would expect there to be significant disk emission in the $K$ band due to
reprocessing of stellar radiation and/or accretion luminosity (\cite{als};
\cite{kh87}).  In the simulations where $R_{\rm in}=R_\star$ our
$K$ light curves are likely to be in error.  For the simulations with larger
$R_{\rm in}$, the star dominates the flux at $K$ and it is valid that we
ignore disk emission.

\section{Results of Models}

We have investigated 27 different system geometries that arise from our
choices of three spot sizes, three spot latitudes, and three values for the
inner disk truncation radius.  For each geometry we have light curves at ten
inclination angles in each of the $U$, $V$, $I$, and $K$ passbands (the fluxes
computed are actually the monochromatic fluxes at the central wavelength of the passband). For the flat
disk simulation we therefore have 1080 simulated light curves.  We have also
investigated
flared disks for a subset of the system geometries in order to ascertain the
effects of a more realistic disk geometry on the light curves. A summary of
the relevant model parameters, with the ranges of values over which we vary
them, is given in Table 2. 

We begin this section by describing the
predominant effects on the photopolarimetry arising from changes in the system
parameters with the flat disk geometry.  We then investigate the flared
disk geometry and report the differences on the light curves for the flat and
flared disks. As might be expected, there is considerable degeneracy
among the parameters in our models; a given set of model parameters is not
generally prescriptive in a quantitative way of any single photopolarimetric
diagnostic we
have studied. Thus, we attempt in this section to
describe those features of photopolarimetric variability
which, if observed, would be most useful, either individually or in
combination, in constraining T Tauri system parameters. We end this section
by briefly exploring the effect on observables of using a ``ring-like"
spot geometry (as in \cite{mahdavi}).

\subsection{Flat Disks}

We have employed a simplistic disk geometry for the bulk of
our modeling for computational expedience. The simple density structure of the
disk (Eq. 1) has the advantage that the optical depth calculation in the
radiation transfer is computed analytically. Unfortunately, this simple
density structure, and in particular our use of a wavelength-independent
midplane optical depth, does possess some undesirable properties. In particular, 
a number of our models show light from the spot on the lower hemisphere
``leaking"
through the optically thin outer disk regions in nearly edge-on viewing
geometries. In edge-on cases, therefore, the flared disk geometry we consider 
in section 3.2 is more realistic. 

We now detail the effect on the light curves of varying
the spot size, latitude, truncation radius, inclination and wavelength. In
Figures 1-3 we present a graphical summary of the predominant effects on
photometric and polarimetric variability as a result of changing the key 
system parameters: 1) spot latitude ($\phi_s$), 2) inner disk truncation radius 
($R_{\rm in}$), 3) inclination ($i$). 

\subsubsection{Spot Size and Latitude}

For a given spot latitude and system inclination, we find that changing the
spot size has only a minor effect on the magnitude of the photometric
variations
(variability increases slightly as $\theta_s$ increases). Our analysis of
analytic inversion techniques (Section 5) confirms that amplitudes of
photometric
variability are weakly dependent on spot size; spot temperature is a far more
deterministic parameter of photometric variability. Thus, the magnitude of
photometric variability is not, by itself, a useful diagnostic for
constraining spot size. However there are cases,
high latitude spots in particular, where $\theta_s$ determines whether the
spot will be completely occulted by the star as it rotates to the back side
(when $\phi_s + \theta_s < i$).  Flat-bottomed light
curves are produced in such cases (Figures 1b, 2a, 3c), in contrast to the 
sinusoidal light curve morphologies produced otherwise. For the range of spot
sizes
considered here, low latitude spots almost invariably produce flat-bottomed
light
curves, except in the rare case of nearly pole-on viewing. Thus, where
amplitudes of photometric variations alone do not uniquely constrain spot
latitude or size, flat-bottomed light curve morphologies do suggest
low latitude spots. This is especially true if spots are believed to have large
areal coverages (via, e.g., analytic modeling; see section 5), since larger
spots must be situated at lower latitudes in order to be fully occulted on the
back side of the star (though existing analytic modeling studies of spots on T 
Tauri stars have shown that hot spots tend to be small).

Despite the photometric variability produced by spots at all latitudes, we
find that high latitude spots
show little ($\sim 0.1\%$) polarimetric variability during the rotation period 
(Figure 1a). This is in contrast to the larger variability ($\sim 1\%$)
seen with low-latitude spots (Figure 1c).
This is because the light from high latitude spots cannot illuminate the disk at
small radii where most of the polarized light originates --- at small radii
the disk is denser and the spot flux is larger, yielding more scattered
(and hence more polarized) light.  This result is contrary to
the notion that hot spots necessarily give rise to rotationally
modulated polarimetric variability (e.g. \cite{gullgahm}). Thus,
in the context of the model studied here, a lack of periodic polarimetric
variability in the presence of periodic photometric variability implies
high latitude spots. As a caveat, however, we find that the presence of a large
inner disk hole will suppress polarimetric variability even if the spots
are at relatively low latitudes. Diagnostics of inner-disk material, such
as $H-K$ excess, can thus aid in constraining spot latitude via the
polarimetric diagnostics suggested here.

An example may further elucidate the above discussion. Consider a system with 
observed periodic photometric variability consistent with hot spots,
which shows infrared signatures of inner disk material (e.g. $H-K$ excess),
but which shows no
periodic polarimetric variability. According to our modeling, such a system
must possess spots at high latitude.
For consistency, the morphology of the photometric light curve should be
sinusoidal for typical spot sizes. A non-sinusoidal (i.e. flat-bottomed)
morphology in this case would imply a very small spot or a very high system
inclination (although flared disks will tend to bias observations against
systems with high inclinations, as discussed below).

\subsubsection{Disk Truncation Radius}

Varying the disk truncation radius has a somewhat curious effect on the
morphology of the photometric light curves. In particular,
as the inner truncation radius is increased, certain spot latitude/system
inclination geometries allow the
spot on the lower hemisphere to be viewed momentarily through the inner disk
hole as the star rotates. Double-peaked light curves result in such cases
(Figures 1c, 2b, and 2c). When these double-peaked photometric morphologies
occur,
we also see double-peaked variability in both the percent polarization and in 
the polarization position angle. Systems displaying such double-peaked
light curve morphologies, especially in systems for which inner disk truncation
radii are known (via, e.g., SED modeling), are narrowly constrained in the
spot latitude/inclination angle parameter subspace; this double-peaked effect 
is not seen for spots at very high latitudes, nor for systems seen at
very high or very low inclination angles.
Thus, though observations of double-peaked light curve morphologies in
systems with known inner disk holes are a potentially powerful means for
constraining spot latitude and system inclination, we do not expect that such
systems will occur with great frequency. This phenomenon is described more
fully in the following section.

Changes in the inner disk truncation radius also have observable effects on
the emergent polarization, with the percent polarization decreasing with
increasing truncation radius. This is because
most scattered light originates close to the star.
Removing this scattering material by increasing the
truncation radius leads to a decrease in the overall level of polarization.
Despite the correlation of disk truncation radius and percent polarization in our 
idealized treatment, we caution that percent polarization is highly
degenerate in other model parameters, and that the actual percent polarization
scale is dependent on the poorly known scattering properties associated
with the disk material.

Though our modeling indicates that photopolarimetry cannot replace
the power of modeling spectral energy distributions (SEDs) as a diagnostic 
of disk truncation radius, our
results do nonetheless provide a testable prediction of magnetospheric
accretion theory in systems with known inner disk holes.
The observation of double-peaked light curves would offer support for the
two-spot accretion geometry suggested by theory. In combination with 
measured SEDs, such observations can be used to constrain other system parameters, 
such as system inclination (see below). 

\subsubsection{System Inclination}

As already stated, the system inclination is a
contributing factor in the magnitude and in the morphology of the photometric
variations. In particular, at higher inclinations the variability
becomes greater because a larger fraction of the visible spot disappears
behind the back side of the star.
The most highly inclined systems are very likely to result in flat-bottomed
light curves.

Perhaps more significantly, the inclination (for favorable spot latitudes
and disk truncation
radii) determines whether the spot on the lower hemisphere may be seen
(double-peaked light curves). Because
the lower hemisphere spot can be seen for a narrow range of inclination
angles (for a given spot latitude), the observation of double-peaked
photometric light curves provides a constraint on the system inclination,
especially if the inner disk truncation radius is known. 
The minimum inclination at which double-peaked light curves occur is simply
that at which the second spot first becomes visible on the limb of the star, 
while
the maximum inclination is that at which the inner edge of the disk obscures
the second spot from view, and is given analytically by
\begin{equation}
i_{\rm max} = \cos^{-1} \left\{ \frac{\sin (\phi_s - \theta_s)}{\left[
1 + \left( \frac{R_{\rm in}}{R_\star} \right)^2 - 2 \left( \frac{R_{\rm
in}}{R_\star}
\right) \cos (\phi_s - \theta_s) \right]^{1/2} } \right\} \; .
\end{equation}
This result suggests that double-peaked light curve morphologies will occur
only very rarely, not only because the range of inclinations which produce
such morphologies is small, but because even in those systems possessing the
requisite inner disk holes the spots must be positioned at
latitudes which are lower than that expected by theory.
As an example, consider
a system with an inner disk radius of $R_{\rm in} = 3 R_\star$, with spots at
(relatively low) latitudes $\phi_s = \pm 65^\circ$, and with angular radii
$\theta_s = 6.6^\circ$. Such a system produces double-peaked light curves
for inclinations in the range $60^\circ < i < 70^\circ$. Double-peaked light 
curves will be seen, therefore, in only $\sim 15\%$ of systems with this particular 
geometry, considering random inclination effects alone. 

In general the percent polarization increases with inclination.
Polarization probes the asymmetry of a system --- pole-on disks are
rotationally symmetric and produce no polarization, while edge-on disks
present a highly asymmetric scattering surface to the observer and produce the largest
polarization levels (e.g., Brown, MacLean, \& Emslie 1978). As with disk truncation
radius, we caution against mapping a particular percent polarization
level to a particular system inclination.

In contrast to the percent polarization, changes in the
polarimetric position angle are most pronounced in systems seen at low
inclination (Figure 3). Scattered light is polarized perpendicular
to the plane of scattering (plane containing the incident and scattered rays).
When the disk is illuminated by a bright spot, the polarization
position angle is dominated by light scattered from this spot-illuminated
area.  As the star rotates, this area moves around the disk and the orientation
of its associated scattering plane changes relative to the observer.
At low inclinations, the orientation of the scattering plane relative to the
observer undergoes large variations.  For more inclined disks, an observer sees
less variation in the orientation of the scattering planes as the
spot-illuminated
area moves around the disk.  Therefore the largest position angle variability
occurs for the pole-on systems where the scattering planes change the most
during a rotation period. Position angle variability can show amplitudes of
$\sim 90^\circ$ for the most pole-on viewing angles. Indeed, periodic
high-amplitude position angle variability is expected in the absence of any
photometric variability in the most extreme case of directly pole-on
inclination.

\subsubsection{Wavelength}

In general, the photometric behavior at the different wavelengths we
studied mimic one another. The largest effect in changing the wavelength is
the increased contrast
between the brightness of the spot and star at shorter wavelengths, since
for the chosen star and spot temperatures, we are on the Wein side of the
blackbody distribution.  This results in the largest photometric variability
occuring at shorter wavelengths.

The blackbody contrast effect that is so prominent in the photometric
variability is mitigated in the polarimetric variability by the strong
wavelength dependence of the
dust scattered polarization and albedo (Table~1).  Although the albedo
decreases toward longer wavelengths (resulting in less scattering), the
increase in dust scattered polarization, $p$, is the dominating factor so that
the polarization increases as we go into the near IR (e.g., Whitney \& Hartmann
1992). Note, however, that the polarization does decrease in the near IR
when we consider a flared disk geometry (see below). 
The blackbody contrast also results in a rotation of the polarization
position angle with wavelength at a given instant.  This arises because
the position angle probes the axisymmetry of a system: an axisymmetric system
would show no variations in position angle (or polarization) with rotational
phase, and position angle variability increases as the system geometry or
illumination departs from being axisymmetric.  For example, at phase 1.0 in
Figure 3a, the polarization position angle in $U$ is near $90^\circ$ while in
$I$ it is $20^\circ$.  This arises because the spot/star luminosity ratio and
hence departure from axisymmetry is much larger at $U$ than at $I$.  See
Section~4 for a discussion of this in relation to observations.

\subsection{Flared Disks}

Given the size of the parameter space that we chose to explore, we ran our
simulations using a flat disk for expediency in the Monte Carlo radiation
transfer.  However, disks around classical T Tauri stars are expected to
have a scale height that increases with distance from the central star 
(e.g. \cite{kh87}).
We therefore investigated the effect of a flared disk geometry on the
photopolarimetric light curves of some of the models. The
photopolarimetric variability is generally qualitatively similar to the
flat disk models (Figure 4).

There are three major differences arising from the dense flared disk.  The
first is
that the denser disk allows for more scattered radiation and hence a higher
overall percent polarization, especially at high inclinations, although 
the morphology of the polarimetric
variations are similar to that observed for the flat disk (Figure 5a). 
As a consequence,
the flared disk allows for considerably larger polarimetric variability than
the flat disk, especially at high inclinations. As an example of this, Figure 5b 
compares the polarization
percent variability, as a function of inclination angle, arising from the flat
and flared disk models used in Figures 1 and 4. The flat disk model can produce
variability in percent polarization of at most 0.3\%, while the flared disk
model can produce variability of 1-5\% at inclinations greater than $70^\circ$.
Secondly, for edge-on viewing the flux is greatly reduced since the
star is viewed through the outer flared regions.  This effect is displayed in
Figure 5c which shows the flux as a function of viewing angle for our flared
disk geometry.
Although the variability of the {\it differential} magnitude may be
substantial,
it is unlikely, given the large circumstellar extinctions, that the star would
be detected at visible wavelengths.  Indeed it is unlikely that such systems
would be classified as classical T-Tauri stars (Whitney \& Hartmann 1992,
1993).
Finally, unlike the monotonic increase of the percent polarization with
wavelength in the flat disk geometry (Figures 1-3), the percent polarization
decreases at near-IR wavelengths with the flared disk geometry (Figure 4).
This is because in our flared disk simulation the disk optical depth decreases
into the IR.  The combination of a lower albedo and optical depth results in
less scattering and offsets the increase in dust scattered polarization with
wavelength (Table~1).  For high inclinations, however, when the star is
occulted by the disk and the light we see is predominantly highly polarized
scattered light the polarization does increase in the IR (see also Whitney \& 
Hartmann 1992, 1993).

\subsection{Ring-Like Spots}

Recent detailed calculations of magnetospheric accretion in classical
T Tauri stars has involved modeling spots as ``rings" or ``annuli"
(\cite{mahdavi}). Though we do not attempt to argue in this study
for one spot geometry over another on a theoretical basis, we do comment on
the effect of a ring-like spot geometry on the photopolarimetric observables
we have modeled.

We generated a single model (with a flared disk), employing typical system
parameters, at one
wavelength (Figure 6), to see if we could detect any significant differences in the
resulting photopolarimetric behavior which might allow one to distinguish
between accretion rings and spots on the basis of photopolarimetry. We 
find no such differences. As with filled-in spots, the
morphology of the photometric variation is sinusoidal, its amplitude
consistent with the areal coverage of the ring (cf. Figure 4a). Our modeling
thus suggests that photopolarimetry may not be useful for distinguishing 
fine details of accretion spot geometry.

\section{Comparison to Observations}

Hot spots have been implicated in the observation of periodic photometric
variability in scores of T Tauri systems (e.g. \cite{bouvier93}). Hot spots 
are also suspected to be the cause of the more
erratic variability seen in the large amplitude irregular variables 
(e.g. Eaton, Herbst, \& Hillenbrand 1995).
This may arise because accretion onto hot spots is likely to be dynamically
active. Thus, morphologies of hot spot light curves are likely to behave more
stochastically than the strictly periodic behavior of our model systems.
Nonetheless, stable periodic variability arising from hot spots has been seen
(\cite{choi}; \cite{herbst94}; \cite{bouvier93}).
Bouvier et al. (1993) find that hot spots are present in the majority of the
classical T Tauri stars in their sample, indicating that hot spots do
sometimes exist stably for timescales long compared to one stellar rotation period.
Bouvier et al. (1993) further find that hot spots typically cover only a few 
percent of the stellar surface, consistent with our use of relatively small
spots in our models.

Observational studies have shown that, almost without exception,
the periodic photometric variations of T Tauri stars have a sinusoidal
morphology.
The morphology of our photometric light curves are generally sinusoidal except
when the spots are occulted (giving flat bottomed light curves) or at low
latitude and the second spot is seen (giving double peaked light curves). 
Bouvier \& Bertout (1989) model one star they observed (DF Tau) with a flat bottomed
light curve. But such light curve morphologies have been seen only rarely.
The observational absence of flat-bottomed morphologies indicates that
spots are at least partially visible throughout the stellar rotation period.
On the basis of our modeling, this suggests that high latitude spots prevail 
among those systems which have been studied, since low latitude spots will
be occulted by the star unless the system is viewed almost pole-on. Our models
further indicate that these systems are predominantly observed at low
inclination, since even at high latitude a spot can be occulted if viewed at
high inclination.
We do note that flat-{\it topped} light curves due to cool spots have been
observed more frequently
(e.g. \cite{bouvier89}), suggesting that cool spots do occur at low latitudes or that
some
systems with cool spots are observed at high inclination. Indeed, it is not
surprising within the context of the magnetospheric accretion paradigm that
only cool spots are observed in stars seen at high inclination; hot
spots will only be observed on stars with circumstellar disks, which
will tend to bias observations against highly inclined systems.

The amplitudes of photometric variability seen in our models, $\Delta I \lesssim 
1$ mag (see Figures 1-3), are generally consistent
with those observed (see \cite{mahdavi} for histograms of photometric variability
in Taurus-Auriga and Orion). 
It should be noted that the samples used in studies of
periodic photometric variability (e.g. \cite{bouvier93}; \cite{herbst94}; \cite{choi})
almost certainly contain a mix of classical and weak-lined T Tauri stars. It is
thus unclear the extent to which observations of periodic photometric variability
are dominated by cool spots, bearing no connection to the magnetic accretion model
studied here. Nonetheless, our models demonstrate that hot spots can account for
the observed range of photometric variability among periodic variables. 
Large amplitude irregular variables have been observed 
with somewhat larger amplitudes of variability, sometimes exceeding 2 mag at $I$
(\cite{choi}). This is larger than that produced in any of our models, indicating
that such large amplitude variations are likely due to larger spot/star temperature 
ratios than simulated in our models (\cite{choi} attribute some of these large
amplitude irregular variations to obscuration by circumstellar material on the
basis of light curve morphology). 

Double-peaked light
curves only occur when both spots can be seen during the rotation period,
arising from a favorable combination of spot latitude and inner
disk hole size.  Systems with known inner holes (as inferred from modeling of
spectral energy distributions) should be monitored for this effect. To date, 
there have been no reports of double-peaked light curves attributable to hot
spots. But if this
effect is present and strong (i.e., second spot as bright as primary) it could
mistakenly result in a period determination that is half of the true rotation period,
as has occurred in JW~191 (\cite{choi}). Although double-peaked light curves
do not guarantee that the second spot is located on the lower hemisphere, such
morphologies are nonetheless predicted by the magnetic accretion model and the
observation of such, especially when the two spots are separated in phase by
close to $180^\circ$, merits special attention.

%
Polarization studies of T Tauri stars have been more carefully directed than
photometric variability studies at classical T Tauri stars. Even so, we caution
that classical T Tauri status does not guarantee the presence of hot spots
(e.g. \cite{bouvier93}). Ideally, a measure of spot temperature (see
Section 5) would accompany future polarimetric variability studies.
The polarization levels seen in our models are consistent with those
observed. Menard \& Bastien (1992) present a distribution of observed linear
polarization levels for 122 stars, and find a most common value of
$\sim 1\%$ with a long tail extending to polarizations as high as 
$\sim 15\%$. Figures
1-3 show representative polarizations of $\sim 1-2\%$, but none as high as the
largest observed. Such large polarization can only be reproduced in our 
models when
we employ a flared disk at relatively high inclination (see Figure 5a). At 
such high inclinations, reddening due to circumstellar dust in the flared disk
becomes non-negligible (Figure 5c), so that our models predict that systems which 
exhibit polarizations larger than $\sim 3\%$ should also be highly reddened (e.g.
HL Tau; Menard \& Bastien 1992).

Polarimetric variability has been observed in many systems, typically at
levels of $\lesssim 0.5\%$, and only occassionally at levels of up
to a few percent (\cite{gullgahm}). The results of our models are consistent
with this (Figures 1-4) but in our models variations in polarization $\gtrsim 1\%$
require flared disks at high inclination (Figure 5b).
Unfortunately, little synoptic
photopolarimetric variability work has been carried out, with most of the
existing observations (e.g. Schulte-Ladbeck 1983; \cite{drissen89}; Menard \& 
Bastien 1992; \cite{gullgahm}) either having sparse time-sampling or being
relatively well sampled but spanning only a few days.
Periodicities have been suggested for only a few systems (see, e.g., 
Menard \& Bastien 1992 and \cite{drissen89}), but not yet solidly
confirmed for any. Menard \& Bastien 
propose several mechanisms for the polarimetric variability they observe,
including
star spots and inhomogeneities in the circumstellar environment, and exclude
spots as the likely source of the polarimetric variability due to this apparent 
absence of periodic modulations and to the lack of periodic photometric
variability on polarimetrically variable stars. It should be emphasized 
that very little
of the existing photopolarimetric monitoring work has been well suited for
finding the 1-10 day periods that typify T Tauri stars (\cite{choi}); the 
interpretation that polarimetric variability is never periodic is far 
from secure.
In order for disk inhomogeneities to produce non-periodic variability the disk 
density structure would have to change on short timescales since any
inhomogeneous disk structure which is stable over a few orbital periods would 
produce periodic variability.
Alternatively, if accretion streams are present (as in the magnetospheric
accretion model), these present a natural system asymmetry and source for
scattered polarization.  If the accretion flow in these streams is time
variable, they may provide the inhomogeneities proposed by Menard \& Bastien. 
This will be investigated in a future paper. 
On the basis of our models, a {\it lack} of
polarimetric variability could simply be the result of high latitude spots.
Indeed, many of the systems observed by Menard \& Bastien (1992) show little to no
polarimetric variability but show substantial (i.e. $\Delta\theta 
> 45^\circ$) 
position angle variations (cf. their Table 4), as seen in our models with high
latitude spots observed at low system inclination (e.g. Figure 3a). 

A common feature among much of the existing photopolarimetric monitoring
studies of T Tauri stars is the stochasticity of the polarimetric variability.
Our model systems display only strictly periodic behavior because
we did not include any stochastic component to our simulations. To explore
the extent to which the magnetic accretion model can help explain systems
displaying stochastic polarimetric behavior, we have generated a model (the same
model as in Figure 3b) in which we allow the spot size to vary randomly over three
stellar rotations so as
to simulate, if only crudely, the effect of variable accretion. The resulting
polarimetric variability is shown in Figure 7a. This type of variability, 
if observed,
would likely be interpreted as random variations. Yet
a clear correlation is present between the 
polarimetric and photometric variability (Figure 7b), in the sense of minimum
polarization at maximum brightness. This type of correlation has been observed
in several systems (see \cite{gullgahm}, but see \cite{drissen89}
figure 2 for a plot of RY Lup similar to our Figure 7b). We do not find any reported clear 
instances of systems in which polarimetric variability in the absence of
correlated photometric variability is observed. Such cases would not 
be explained
by our models, and would require mechanisms other than those investigated in
our models, such as time-variable inhomogeneities in the circumstellar
environment (as suggested above, the accretion streams
inherent to the magnetic accretion model might provide a natural explanation 
for these systems).

Our models predict that the percent polarization increases with wavelength
(though our flared disk simulations do show a polarization turnover at $K$).
This
at first appears contrary to observations which typically show a wavelength
dependence of the polarization that peaks aroung the $V$ band
(e.g. \cite{gullgahm}).  However, these observations have not been
corrected for the effects of interstellar polarization.  The wavelength
dependence of interstellar polarization follows a ``Serkowski Law"
(\cite{serkowski}), which peaks at visual wavelengths. 
Since previous studies have focussed on differential polarimetry, the
intrinsic polarization spectrum of these stars is undetermined.  A careful
analysis of spectropolarimetric data with accurate interstellar removal is
therefore required before we can directly compare our wavelength dependent
polarization model predictions with observations.
Gullbring \& Gahm (1996) have noted that the measured position angle for V1121
Oph at a given time is not independent of wavelength, with a maximum rotation
from $U$ to $I$ of $43^\circ$. They attribute this to a wavelength-dependent 
interstellar polarization component. But this behavior is consistent with many 
of our model 
systems in which differences of $\sim 45^\circ$ in position angle between $U$ 
and $I$ are seen at certain phases. To be entirely consistent with our models, 
observed differences in position angles at different wavelengths should vary 
as a function of phase (e.g. Figure 3a).

We have not included in our models cool spots, which are often the
primary source of periodic
photometric variability in classical T Tauri stars (\cite{bouvier93}).
As such, our model predictions of photometric variability only apply to cases
where hot spots are present and dominant. Our model predictions
of polarimetric
variability, however, are not likely to be severely affected by our
omission of cool spots. Hot spots, which in our models have luminosities
comparable to stellar, produce polarimetric variability because of their
large illuminating assymetry. Cool spots, which are typically
only a few hundred degrees cooler than photospheric (\cite{bouvier93}),
do not present such a large asymmetry and will consequently result in smaller
polarization levels and polarimetric variability.

\section{Analytic Inversion}

Analytic inversion techniques are commonly employed to determine spot
properties such as temperature and areal coverage. In its most simple
incarnation, this technique basically consists of a two-component blackbody
fit to the observed {\it amplitudes} of periodic photometric variability at
multiple wavelengths.
Usually, the stellar temperature is known (through the stellar spectral type),
minimizing the number of free parameters to two: spot temperature and size.
Although this type of analytic modeling has been applied by various authors
throughout the literature, no analyses of the reliability of such modeling
have been reported (but see the statistical analysis of Levenberg-Marquardt
techniques by \cite{kovari}). Typically, spot parameters reported
on the basis of analytic inversion techniques are reported with
uncertainties of $\sim 100$K in temperature and a few percent in areal
coverage.

We have applied an analytic inversion
method to our simulated light curves in order to ascertain the reliability
of such modeling in accurately recovering spot properties.
For specificity, we adopt a model similar to that employed by Bouvier et al.
(1993) and Vrba et al. (1988) in which
the wavelength-dependent amplitude of photometric variation, $\Delta
m(\lambda)$,
is parametrized in terms of the spot-to-star flux ratio, $Q(\lambda)$, and 
the projected minimum and maximum spot areal coverage, $f_1$ and $f_2$:
\begin{equation}
{\Delta m(\lambda) = -2.5 \log \left[{{1-\left(1-Q(\lambda)\right) f_2}
\over {1-\left(1-Q(\lambda)\right) f_1}}\right]} \; .
\end{equation}
Unlike these authors, we do not consider the effects of limb darkening, nor
a radiation distribution more sophisticated than a simple blackbody, as
our current Monte Carlo simulation does not employ model atmospheres.

We find that this simple model is able to accurately reproduce the known spot
temperature and size in each of our Monte Carlo simulations. This is perhaps
not surprising, considering
that the simulations consist --- as does the analytic model --- of two
simple blackbody components,
and as our simulated light curves are noiseless. But to what extent can
spot parameters be constrained when real (i.e. noisy) data are modeled? In
order to quantify this, we investigated the dependence of the size of the
solution space as a function of noise in the input data. We generated 1000
$\{\Delta U$,
$\Delta V$, $\Delta I$, $\Delta K\}$ sets for each of our models by adding
normally
distributed noise with standard deviations of $0.01$ mag, $0.03$ mag, and
$0.05$
to the known $\Delta m$. We then computed best-fit values of $T_s$, $f_1$, and
$f_2$ via equation 6 above for each $\{\Delta U$, $\Delta V$, $\Delta I$,
$\Delta
K\}$ set using a chi-square minimization technique,
and take the region of the parameter space containing 95\% of these
best-fit values as the region of 95\% confidence (\cite{nrecipes}).

We find that
the spot temperature is significantly more sensitive to noise in the
photometric
amplitudes than previous use of this type of analysis suggest. Even for 
modest uncertainties in the data
$(\sigma_{\Delta m} = 0.01)$, spot temperatures between 7500K and 15000K yield
strong goodness-of-fit for the particular case shown in Figure 8c. In contrast, 
spot sizes only vary between 3\% and 12\% in this case,
small by most standards. This suggests that error estimates of spot
temperatures commonly quoted in the literature  where this type of
analytic inversion is used (e.g. \cite{bouvier93}) may be underestimated. 
We note that this basic
conclusion is similar to the results of K\H{o}v\'{a}ri \& Bartus (1997), who find
that photometric uncertainties of order $1\%$ are sufficient to make their
spot parameter determination become unstable. We further find that the
largest density of solutions for photometric amplitude uncertainties of
$\gtrsim 3\%$ occurs at spot temperatures which are cooler --- and spot sizes
which are larger --- than the true values. This feature of the solution space
may serve to partially compensate for the tendency to underestimate spot sizes 
due to projection effects (\cite{bouvier93}).

Despite this difficulty in accurately determining spot temperatures given even
small formal uncertainties in photometric amplitudes, we find that it is
relatively straightforward to distinguish hot spots from cool ones. In fact,
we find that it is nearly impossible to force the analytic inversion studied
here to incorrectly label a hot spot as a cool one, assuming ``reasonable"
uncertainties in the photometric data (pathological solutions --- e.g. a cool ``spot"
covering 95\% of the surface --- excluded). For the case considered in Figure 8,
cool spots are not inferred with any frequency until noise of order 0.20 mag
is added to the photometric amplitudes.

\section{Summary and Conclusions}

We have modeled the photopolarimetric variations arising from a classical T
Tauri
star/disk system possessing hot spots on the stellar surface in an attempt to
identify useful observational diagnostics of system parameters of interest 
and to assess the ability of the magnetic accretion model to explain existing
observations of photopolarimetric variability in T Tauri stars.
Though we have restricted our attention in this initial study to a simple
interpretation of
magnetospheric accretion, the Monte Carlo technique we employ can be applied to
any other model of interest (e.g., flared disks, hot ``rings", cool spots,
stellar atmospheres). We will
investigate the effects of the actual magnetospheric accretion flow on observed 
polarimetric variability in a future study.

On the basis of our modeling, we confirm that the magnetic accretion model
is consistent with existing observations of photopolarimetric variability
in T Tauri stars, and have identified
the following observational diagnostics of key system parameters. These
serve as testable predictions of the magnetospheric accretion model, and
as a guide to future photopolarimetric monitoring studies of T Tauri stars:
\begin{itemize}
\item Double peaked light curves --- reveal existence of spot on lower
hemisphere.
This effect requires a favorable combination of spot latitude, system
inclination,
and inner disk truncation radius. In particular, if a ``typical" disk
truncation
radius of $R_{\rm in} = 3 R_\star$ is assumed, spots must be situated at low
latitudes ($\phi_s \lesssim 60^\circ$) and
the system inclination is somewhat constrained ($\phi_s \lesssim i \lesssim
75^\circ$).
Given the theoretical bias against such low-latitude spots, and the small
region
of parameter space occupied by the appropriate combination of system
parameters,
we do not expect that such double-peaked light curve morphologies will occur
with great frequency. We note however that Choi \& Herbst (1996) have observed
``period-doubling" in JW191, presumably the result of a second spot.

\item Flat-bottomed light curves (spot is completely occulted by star) ---
imply low spot
latitude.  For low inclinations, a flat-bottomed morphology could imply a high 
latitude spot of very small size. In these cases,
analytic inversion techniques such as that employed by Bouvier et al. (1993)
can
provide an independent estimate of spot size. The general lack of this morphology among
hundreds of systems observed suggests that low latitude spots and high
inclination rarely if ever occur among classical T Tauri stars.

\item Large polarization position angle variations --- strongly indicative of low inclination.
Figure 3 shows how amplitude of position angle variability is related to
inclination. The effect is most prominent in $U$ band where the spot radiates 
most efficiently relative to the photosphere.

\item Large percent polarization variations ($\gtrsim 1\%$) and/or high
polarization values ($\gtrsim 3\%$) --- require flared disk geometry. Systems
exhibiting very large polarizations will likely be highly reddened by
circumstellar dust in the flared disk.

\item Lack of variability in linear polarization (in the presence of
photometric variability) --- indicative of high latitude
spots. Figure 1 shows how a spot situated at high latitude does not produce
significant polarization variability despite significant photometric
variability.
This is contrary to the periodic variability commonly expected in polarization
variability studies. Hot spots have been mistakenly discounted in systems where
no polarimetric variability has been observed (e.g. Menard \& Bastien 1992; \cite{gullgahm}).
If high latitude spots are common among classical T Tauri stars, we expect that 
such behavior will typify observations of these objects. Such systems observed
at low inclination will display large position angle variations despite showing no
discernible variability in linear polarization. Numerous such cases have been observed
(see Menard \& Bastien 1992).
Hot spots have also been discounted in systems in which polarimetric variability 
is highly stochastic. In this case, hot spots may also be responsible if 
anti-correlated photometric variability is observed (in the sense of minimum polarization
at maximum brightness).
\end{itemize}

We find that wavelength-dependent polarization position angles arise naturally
in our models, and do not require the invocation of multiple interstellar
polarization components.

We find that disk truncation radius is not prescriptive of
either the photometric or polarimetric observational diagnostics considered
in this study. However, cleared inner disk regions are required if the double-peaked 
light curve morphology discussed above is to be observed.

These results are not significantly altered when we model accretion spots as rings.

Lastly, we confirm the applicability of analyic inversion techniques for
recovering spot properties, such as size and temperature, from observations
of periodic photometric variability at multiple wavelengths. We find that
observational uncertainties
will tend to cause such techniques to err more strongly in determining spot
temperatures than has been suggested by some reports in the literature 
employing this type of analysis. We also find that
solutions tend to systematically underestimate spot temperatures and to
overestimate spot sizes. To the extent
that one is interested simply in determining whether spots are hotter or
cooler than stellar photospheres,
we find that a technique similar to that employed by Bouvier et al. (1993)
and Vrba et al. (1988) does so reliably. This result is encouraging,
considering
studies of periodic photometric variability in young stars are often concerned with 
simply determining whether spots are hot or cool.

\acknowledgements

We thank Scott Kenyon, Robert Mathieu, Barbara Whitney, David Cohen, and
Gina Brissenden for
useful discussions. This paper was improved considerably by criticisms from
an anonymous referee. This work has been funded through an NSF graduate student
fellowship and a grant from NASA's Long Term Space Astrophysics Research
Program (NAG5-6039).  The radiation transfer models were run on the SGI 
cluster at the University of Wisconsin Astronomy Department.

\clearpage
\figcaption{The variations in the photometry, percent polarization, and
polarization position angle are shown over one rotation period for three
different choices of spot latitude: a) $\phi_s = 80^\circ$, b) $\phi_s =
65^\circ$,
c) $\phi_s = 45^\circ$. All other model parameters are fixed: $\theta_s =
20^\circ$, $R_{in} = 3 R_\star$, and $i = 63^\circ$. In each panel, the line
style indicates the four different passbands as follows: $U$ (solid line),
$V$ (dotted line), $I$ (dashed line), and $K$ (dotted-dashed line). Note the
progression in light curve morphology from sinusoidal, to nearly 
flat-bottomed (see Figures 2a and 3c for better examples of this), to
double-peaked, as the spot latitude is varied. Note also the lack of
variability
in the percent polarization for the system with the high-latitude spots.}

\figcaption{The variations in the photometry, percent polarization, and
polarization position angle are shown over one rotation period for three
different choices of inner disk truncation radius: a) $R_{in} = R_\star$, b)
$R_{in} = 3 R_\star$, c) $R_{in} = 15 R_\star$. All other model parameters are
fixed: $\theta_s = 11.5^\circ$, $\phi_s = 45^\circ$, and $i = 63^\circ$. 
Line styles are as in Figure 1. Note the transition from flat-bottomed to
double-peaked light curve morphology as the inner disk recedes from the stellar
surface. At $3 R_\star$, the spot on the lower hemisphere is fully revealed;
the only change in increasing the disk truncation radius further is to decrease
the amount of polarized radiation. Note that the small hump seen in the $U$
light curve in panel a) at
phase $0.5$ is a spurious feature resulting from light from the spot on the
lower hemisphere leaking through the optically thin outer disk regions (see
text).}

\figcaption{The variations in the photometry, percent polarization, and
polarization position angle are shown over one rotation period for three
different choices of inclination angle: a) $i = 32^\circ$, b) $i = 63^\circ$,
c) $i = 87^\circ$. All other model parameters are fixed: $\theta_s = 20^\circ$,
$\phi_s = 65^\circ$, and $R_{in} = R_\star$. Line styles are as in Figure 1.
Note the progression in the amplitude of the polarization position angle
variations as the system inclination is varied from pole-on to edge-on
viewing. Note again the spurious feature in the $U$ light curve in panel c).}

\figcaption{Same as Figure 1, except with flared disk geometry. Note
the change in polarimetric behavior in $K$.}

\figcaption{The salient differences between flared and flat disk geometries. 
a) The maximum polarization (at $V$) as a function of inclination angle for
a model with a flared disk (solid line) and the same model with a flat disk
(dashed line). Model parameters are as in Figure 1b.
b) Same as a), except polarization variability is shown.
c) The fraction of emergent light relative to the intrinsic
stellar output is shown as a function of system inclincation for the flared
disk geometry. For inclinations $i > 70^\circ$ the outer regions of
the flared disk intercept an increasingly larger fraction of stellar
photons. The emergent fraction exceeds unity for $i < 70^\circ$ because the
disk scatters additional light to the observer.}

\figcaption{Same as Figure 4a (V band only), except with ring-like spots.
The rings have an outer radius of $23^\circ$ and an inner radius of $10^\circ$,
giving them the same areal coverage as the spots in Figure 4a.}

\figcaption{The time-variable photopolarimetry arising from a model (parameters
same as in Figure 3b, $I$ band) in which variable (random) accretion is simulated
over three stellar rotations.
a) The percent polarization. b) Though individually 
stochastic, the photometric and polarimetric variability are correlated in the
sense of minimum polarization at maximum brightness.}

\figcaption{The $T_s$ vs. $f_2$ parameter subspace is shown for analytic fits
to the photometric amplitudes of the model with $\phi_s = 65^\circ$,
$\theta_s = 20^\circ$, $R_{\rm in} = 3 R_\star$, and $i = 32^\circ$. The true
spot parameters are $T_s = 10000$K, $f_2 = 0.06$. Shown are the best-fit
spot parameters to 1000 realizations of the photometric amplitudes with (a)
0.05 mag noise, (b) 0.03 mag noise, and (c) 0.01 mag noise. The vertical
lines contain 95\% of the points about the true temperature. Note
the asymptotic behavior at low values of the spot temperature as the
stellar temperature ($T_\star = 4000K$) is approached, and the high density of
points at spot temperatures/sizes which are cooler/larger than the actual
values (see text).}

\clearpage
\begin{deluxetable}{lcccc}
\tablecaption{Dust Parameters}
\tablehead{ \colhead {} & \colhead{$\kappa$} & \colhead {$a$} &
\colhead{$g$} & \colhead{$p$} }
\startdata
$U$ \dotfill & 360 & 0.54 & 0.48 & 0.26\nl
$V$ \dotfill & 219 & 0.54 & 0.44 & 0.43\nl
$I$ \dotfill & 105 & 0.49 & 0.29 & 0.70\nl
$K$ \dotfill & 20 & 0.21 & 0.02 & 0.93\nl
\enddata
\tablecomments{Based on the Mathis et al. (1977) grain size distribution
with optical constants from Draine (1985), Drane \& Lee (1984), and Draine
\& Lee (1985). Polarization values are from White (1979).}
\end{deluxetable}
\clearpage

\clearpage
\begin{deluxetable}{cr}
\tablecaption{Model Parameters}
\tablehead{ \colhead {Parameter} & \colhead{Values} }
\startdata
$T_*$ & 4000 K\nl
$T_s$ & 10000 K\nl
$\lambda$ & $U$, $V$, $I$, $K$\nl
$\phi_s$ & $\pm 45^\circ$, $65^\circ$, $80^\circ$\nl
$\theta_s$ & $6.6^\circ$, $11.5^\circ$, $20^\circ$\nl
$i$ & $18^\circ$, $32^\circ$, $41^\circ$, $50^\circ$, $57^\circ$, $63^\circ$, 
$70^\circ$, $76^\circ$, $81^\circ$, $87^\circ$\nl
$R_{\rm in}$ & $R_*$, $3R_*$, $15R_*$\nl
\enddata
\end{deluxetable}
\clearpage

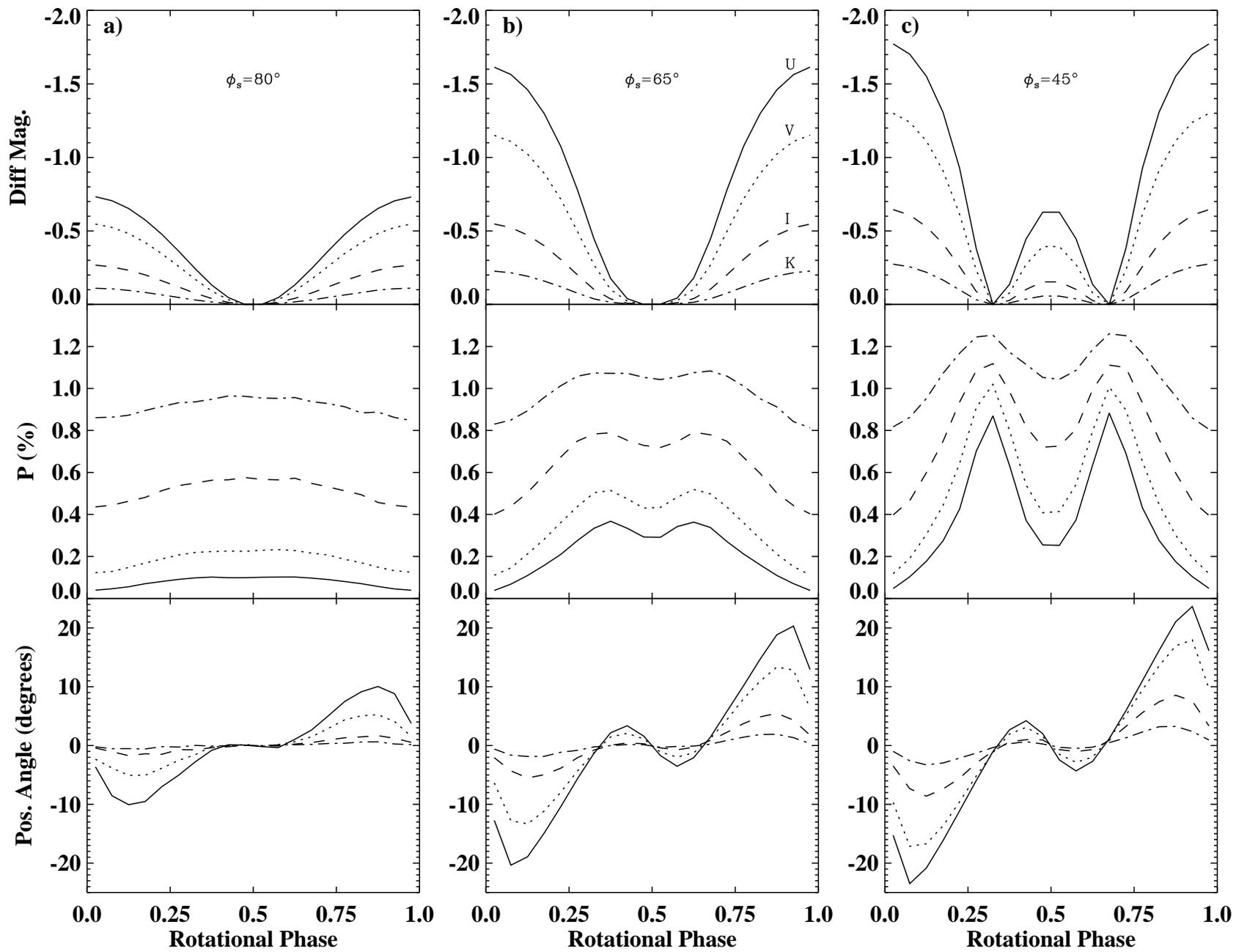
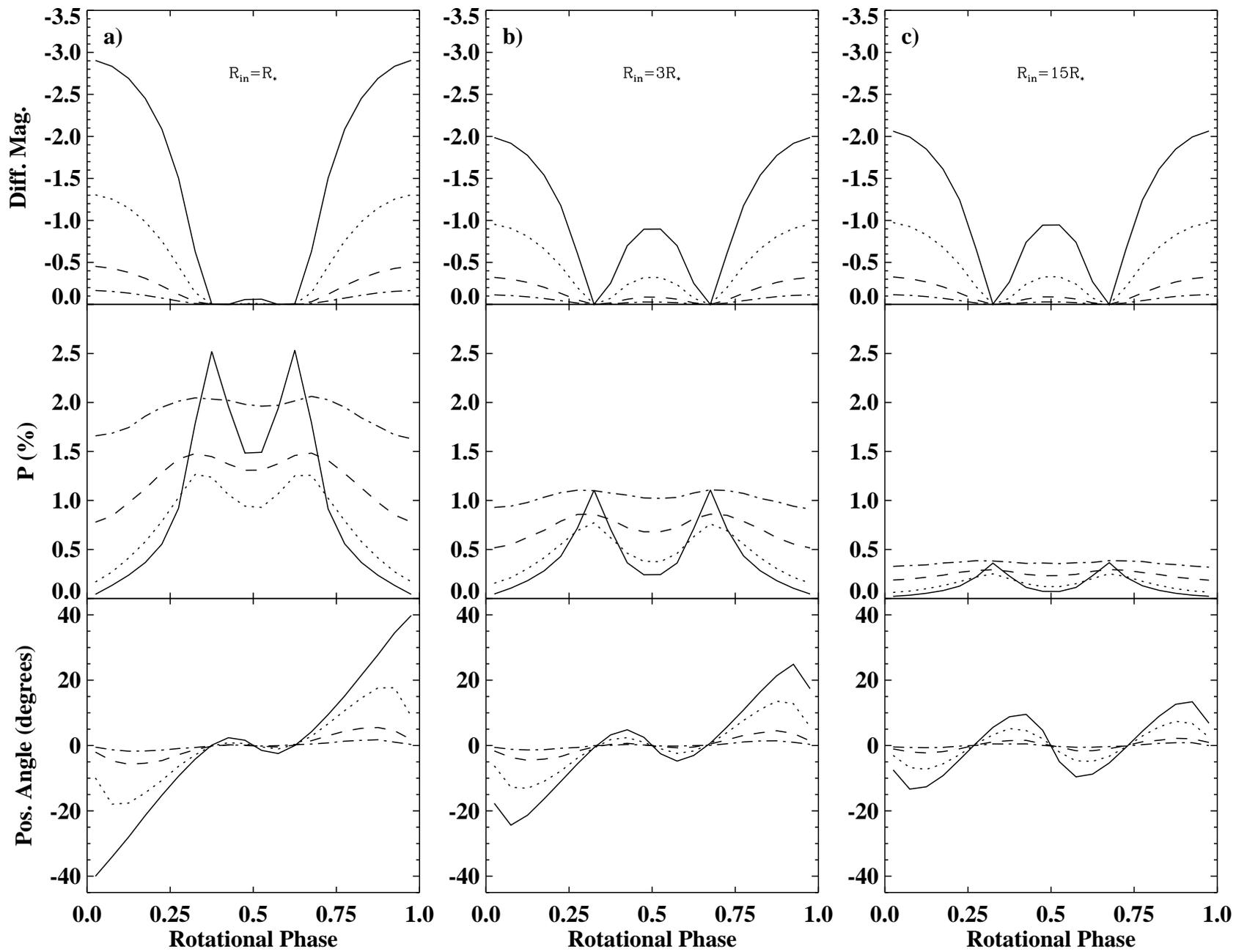
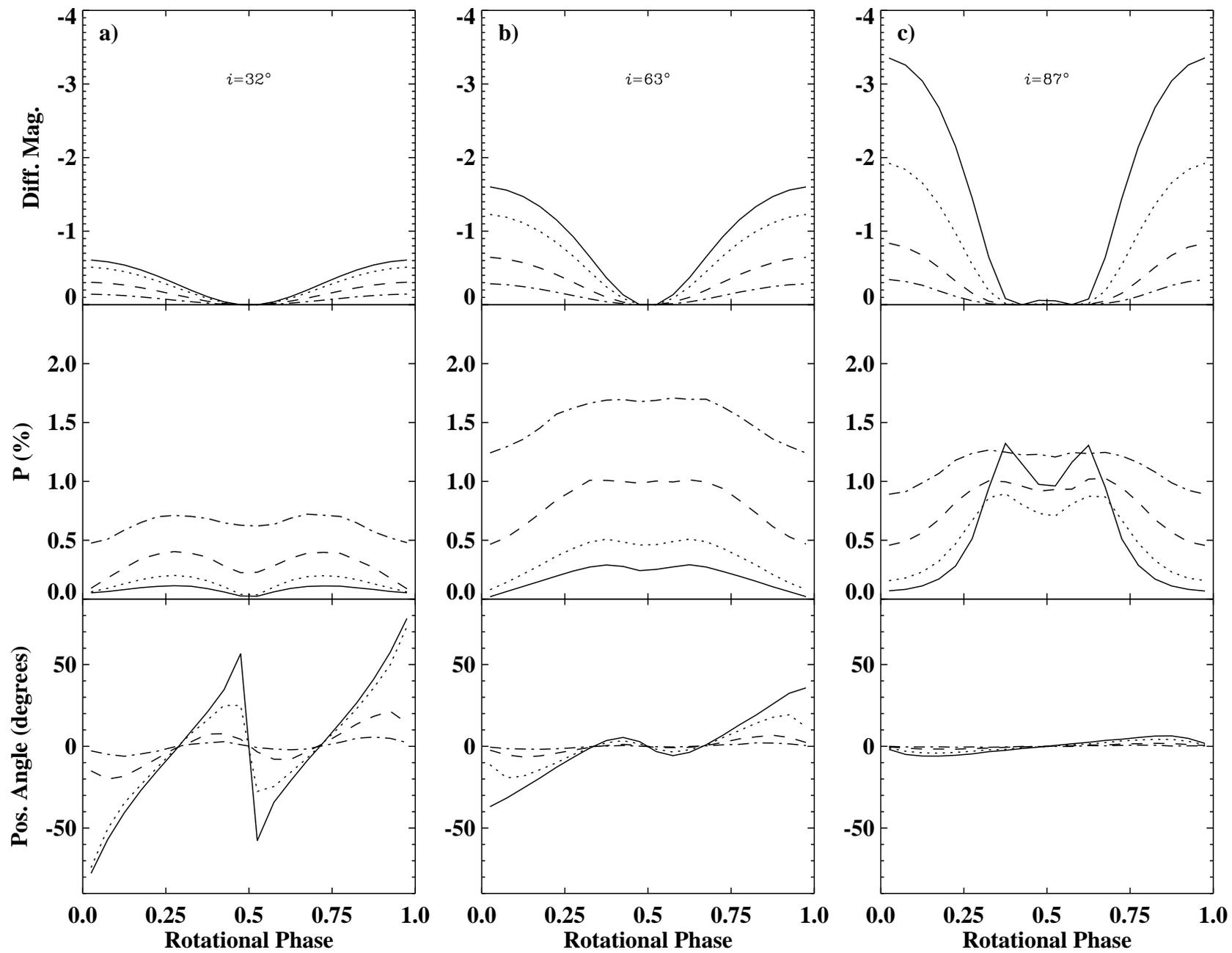
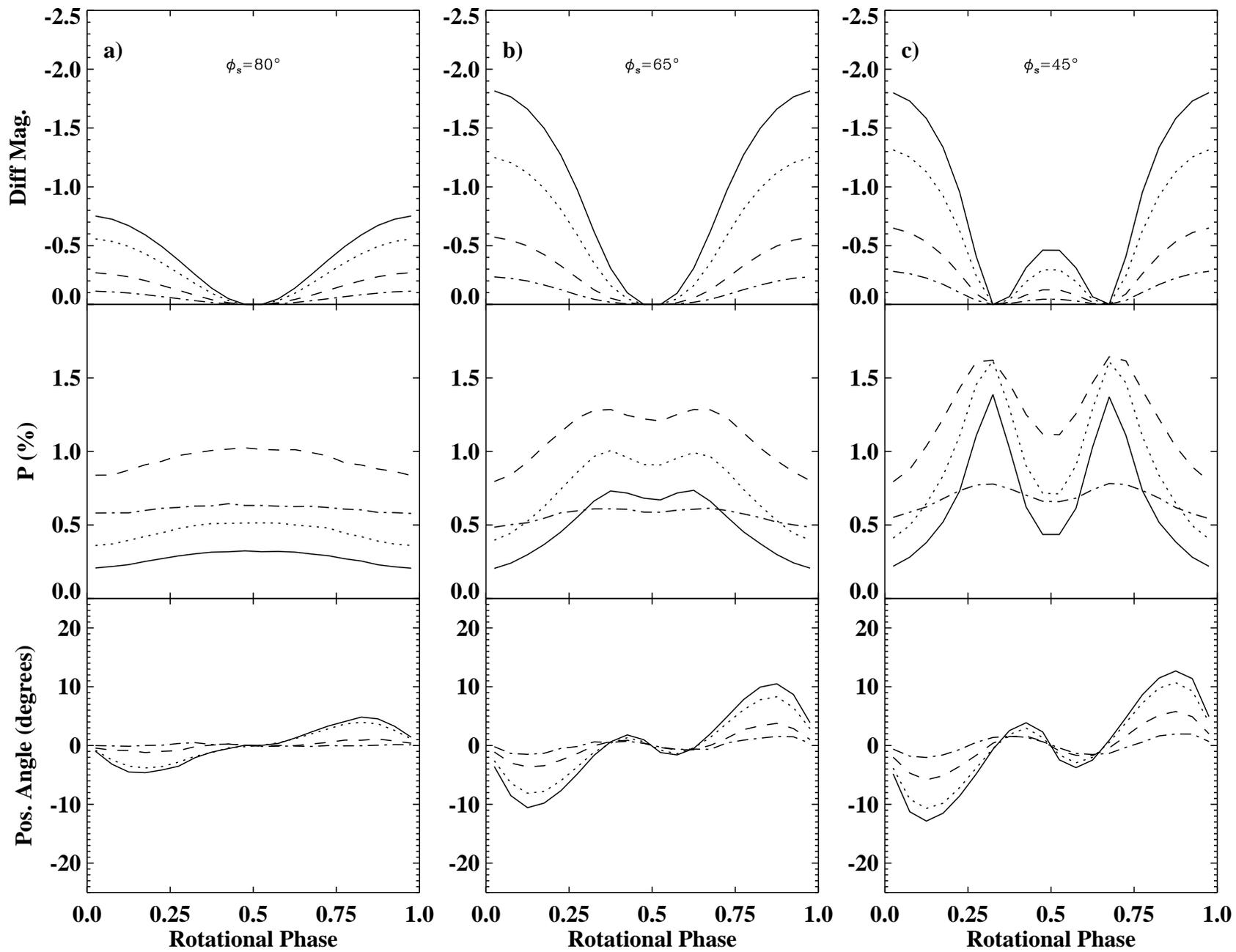
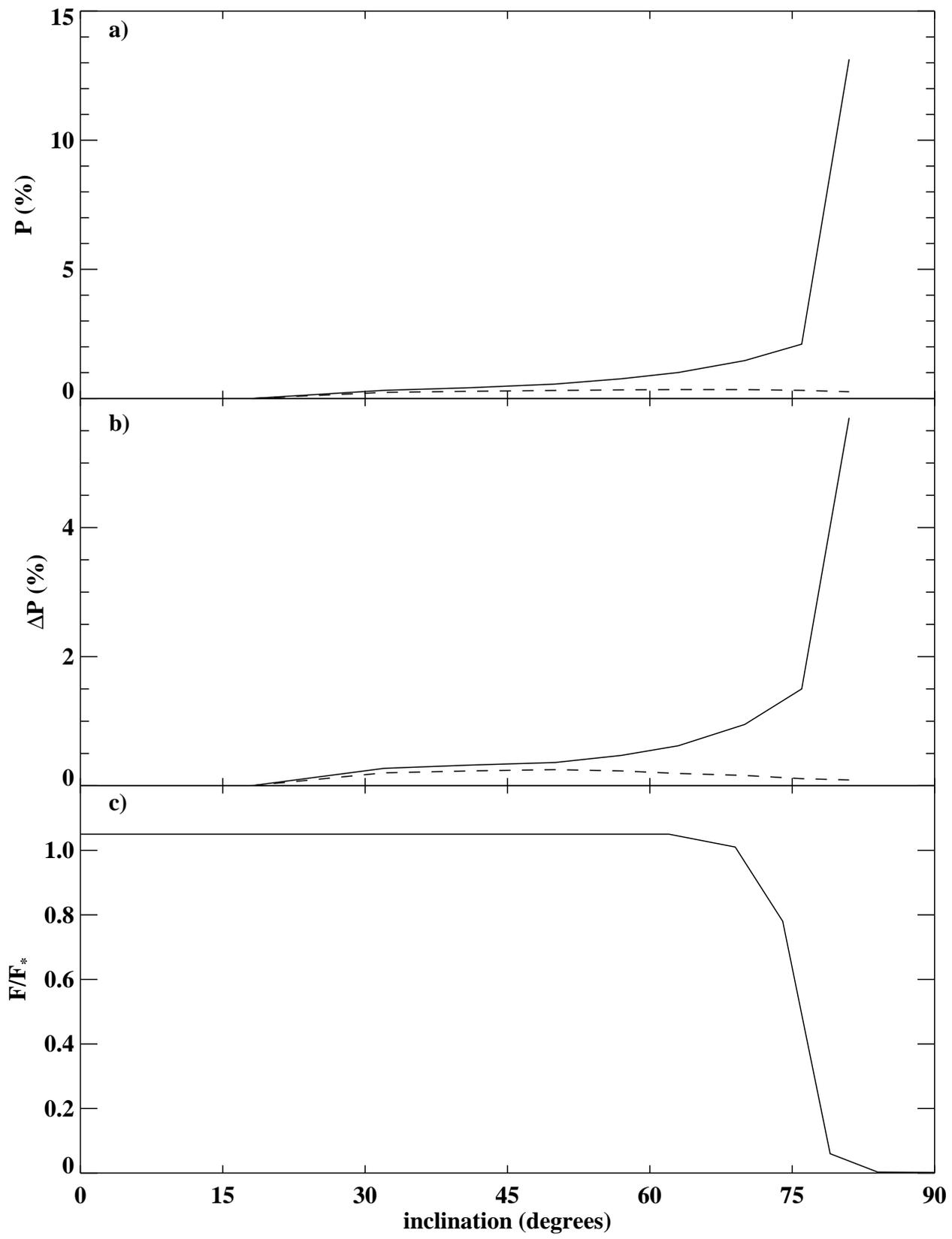
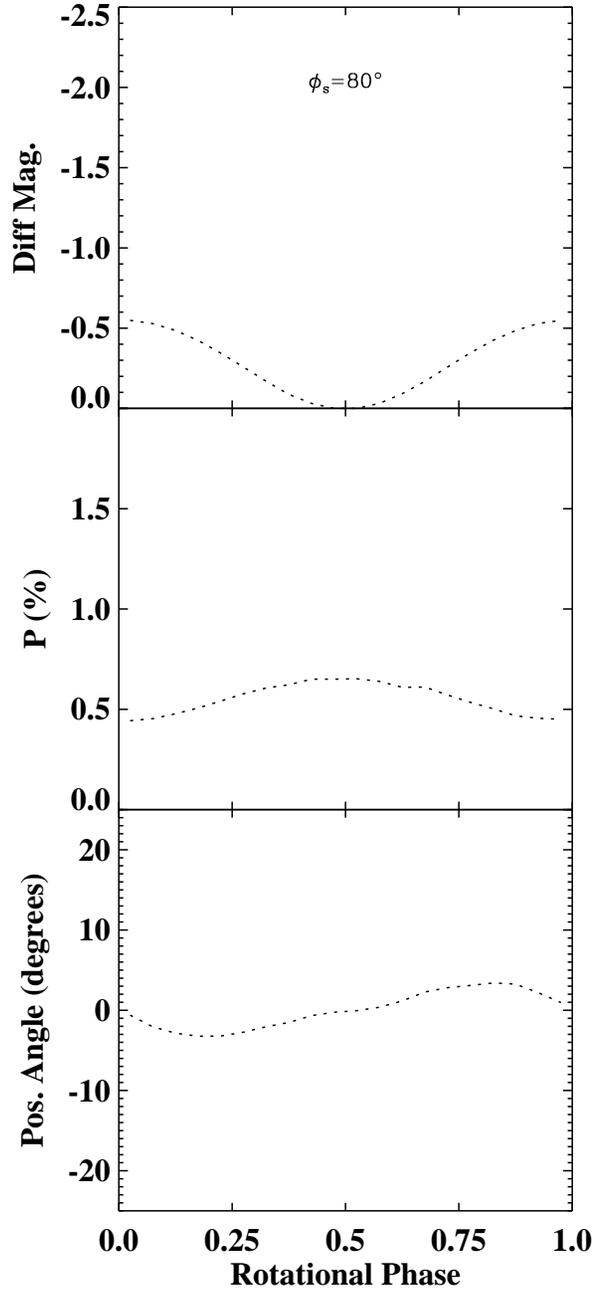
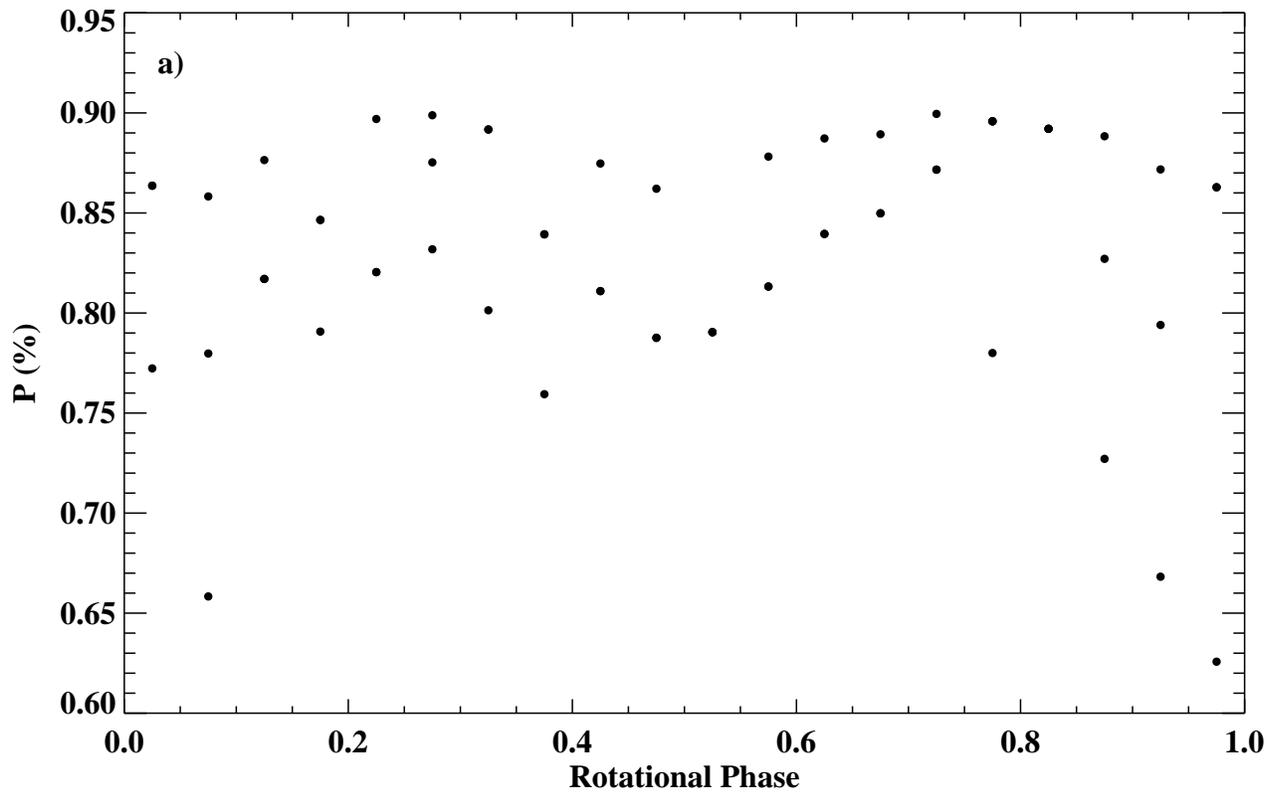
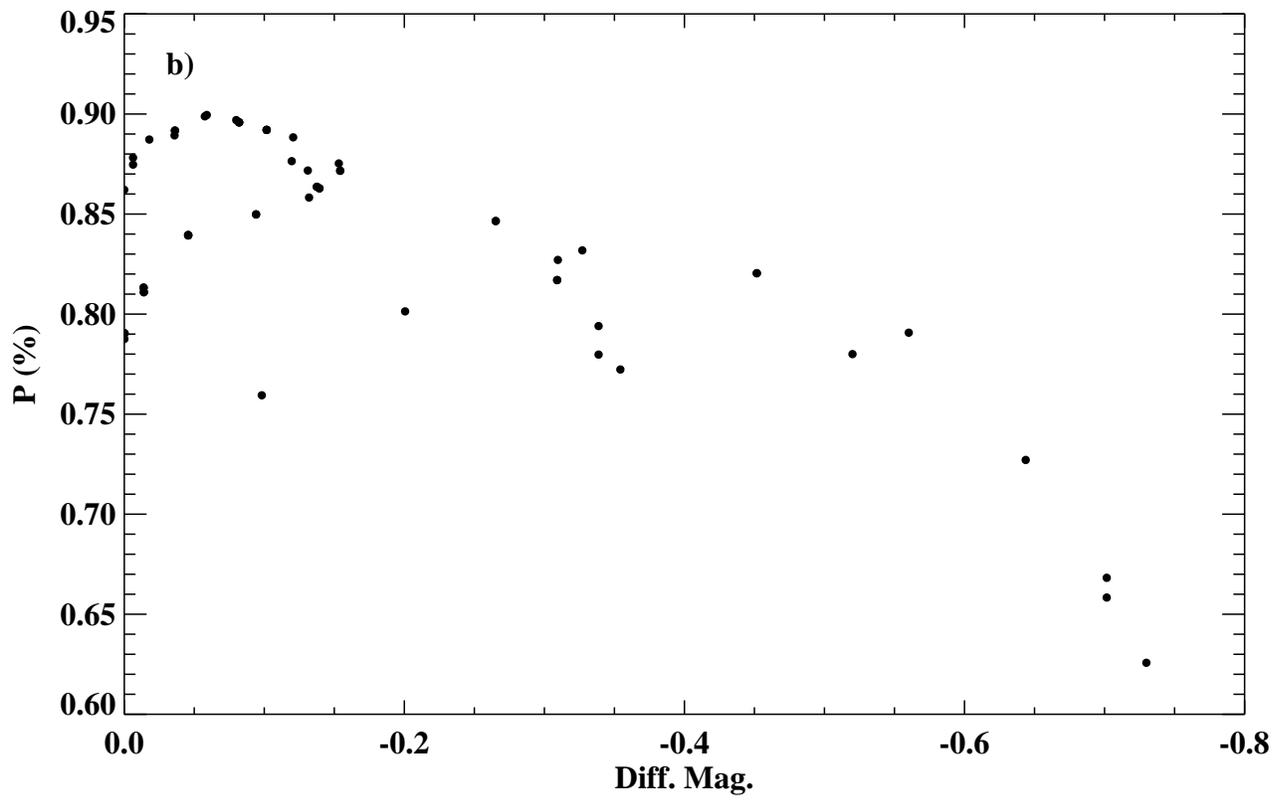
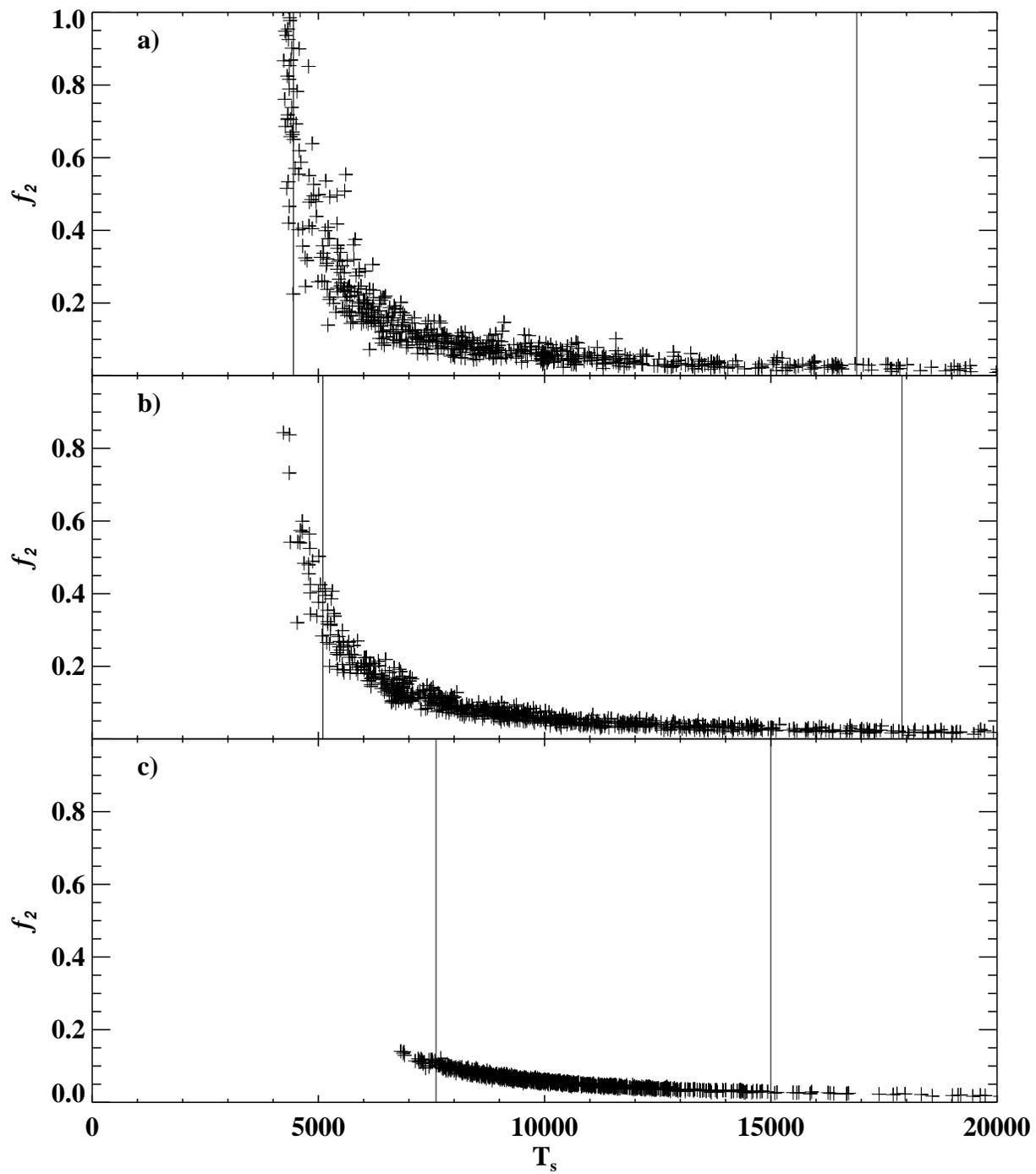
\end{document}